\documentclass[11pt,a4paper]{article}
\usepackage{graphicx,cite}
\usepackage{epsfig}
\usepackage[english]{babel}
\usepackage{amsfonts,amsmath,enumerate,amssymb,mathrsfs,xcolor}
\usepackage{amscd}
\usepackage{color}
\usepackage{hyperref}
\usepackage{theorem}
\DeclareSymbolFont{bbold}{U}{bbold}{m}{n}
\DeclareSymbolFontAlphabet{\mathbbold}{bbold}

\title{}
\author{}
\date{\today}
\newtheorem{definition}{Definition }[section]
\newtheorem{proposition}[definition]{Proposition }
\newtheorem{lemma}[definition]{Lemma }
\newtheorem{theorem}[definition]{Theorem }
\newtheorem{corollary}[definition]{Corollary }
\newtheorem{remark}[definition]{Remark }
\newtheorem{assumption}[definition]{Assumption}

\newtheorem{example}[definition]{Example}
\theorembodyfont{\upshape\it}
\newcommand {\bproof} {\noindent {\bf Proof. }}
\newcommand {\finproof} {\hfill $\Box$ \vskip 5 pt }
\def\indic{1\hspace{-2.5pt}\mathrm{l}}

\def \N{\mathbb{N}}
\def \R{\mathbb{R}}

\def \E{\mathbb{E}}
\def \F{\mathbb{F}}
\def \G{\mathbb{G}}
\def \H{\mathbb{H}}

\def \P{\mathbb{P}}

\def \Ac{{\cal A}}

\def \Ec{{\cal E}}
\def \Fc{{\cal F}}

\def \Pc{{\cal P}}

\def \ni{\noindent}

\def \ep{\hbox{ }\hfill$\Box$}

\def\reff#1{{\rm(\ref{#1})}}

\def\beqs{\begin{eqnarray*}}
\def\enqs{\end{eqnarray*}}
\def\beq{\begin{eqnarray}}
\def\enq{\end{eqnarray}}

\begin{document}
\title{Information uncertainty related to marked random times and optimal investment}
\author{Ying Jiao\thanks{Universit\' e Claude Bernard - Lyon 1, Laboratoire SAF, 50 Avenue Tony Garnier, 69007 Lyon, France. Email:
ying.jiao@univ-lyon1.fr} \qquad  Idris Kharroubi\thanks{Universit\'e Paris Dauphine, , PSL Research University, CNRS UMR 7534, CEREMADE, Place du Mar\'echal De Lattre de Tassigny, 75775 Paris Cedex 16, France. Email: kharroubi@ceremade.dauphine.fr } }
\maketitle
\begin{abstract}
We study an optimal investment problem under default risk where related information such as  loss or recovery at default is considered as an exogenous random mark added at default time. Two types of agents who have different levels of information are considered. We first make precise the insider's information flow by using the theory of enlargement of filtrations and then obtain explicit logarithmic utility maximization results  to compare optimal wealth for the insider and the ordinary agent. 

\vspace{2mm}

\vspace{1mm}
\end{abstract}
%
%

%

\numberwithin{equation}{section}


\section{Introduction}

The optimization problem in presence of uncertainty on a random time is an important subject in finance and insurance, notably for risk and asset management when it concerns a default event or a catastrophe occurrence. Another related source of risk  is the information associated to the random time  concerning  resulting payments, the price impact, the loss given default or the recovery rate etc. Measuring these random quantities is in general difficult since the relevant information on the underlying firm is often not accessible to investors on the market. For example, in the credit risk analysis, modelling the recovery rate is a subtle task (see e.g. Duffie and Singleton \cite[Section 6]{DS2003},  Bakshi et al. \cite{BMZ2006} and Guo et al. \cite{GJZ09}). 
In this paper, we study the optimal investment problem with a random time and consider the information revealed at the random time as an exogenous factor of risk. We suppose that all investors on the market can observe the arrival of the random time such as the occurrence of a default event. However, for the associated information such as the recovery rate, there are two types of investors: the first one is an informed insider  and the second one is an ordinary investor. For example, the insider has private information on the loss or recovery value of a distressed firm at the default time and the ordinary investor has to wait for the legitimate procedure to be finished to know the result.  Both investors aim at maximizing the expected utility on the terminal wealth and each of them will determine the investment strategy based on the corresponding information set. Following Amendinger et al. \cite{AIS98, ABS2003}, we will compare the optimization results and deduce the additional gain of the insider. 

Let the financial market be described by a probability space $(\Omega,\mathcal A,\mathbb P)$ equipped with a reference filtration $\mathbb F=(\mathcal F_t)_{t\geq 0}$ which satisfies the usual conditions. In the literature, the theory of enlargements of filtrations provides essential tools for the modelling of different information flows. In general, the observation of a random time, in particular a default time, is modelled by the progressive enlargement of filtration, as proposed by Elliott et al. \cite{EJY2000} and Bielecki and Rutkowski \cite{BR2002}.  The knowledge of insider information is usually studied by using the initial enlargement of filtration as in \cite{AIS98, ABS2003} and Grorud and Pontier \cite{GP98}. In this paper, we suppose that the filtration $\mathbb F$ represents the market information known by all investors including the default information. Let $\tau$ be an $\mathbb F$-stopping time which represents the default time.  
The information flow associated to $\tau$ is modelled by a random variable $G$ on $(\Omega,\mathcal A)$ valued in a measurable space $(E,\mathcal E)$. In the classic setting of insider information,  $G$ is added to $\mathbb F$ at the initial time  $t=0$, while in our model, the information is added punctually at the random time $\tau$. Therefore, we need to specify the corresponding filtration which is a mixture of the initial and the progressive enlargements.  Let the insider's filtration $\mathbb G=(\mathcal G_t)_{t\geq 0}$ be a punctual enlargement of $\mathbb F$  by adding the information of $G$ at the random time $\tau$. In other words, $\mathbb G$ is the smallest filtration which contains  $\mathbb F$ and such that the random variable $G$ is $\mathcal G_\tau$-measurable. We shall make precise the adapted and predictable processes in the filtration $\mathbb G$ in order to describe investment strategy and wealth processes. As usual, we suppose the density hypothesis of Jacod \cite{Jacod} that the $\mathbb F$-conditional law of $G$ admits a density with respect to its probability law. By adapting arguments in F\"ollmer and Imkeller \cite{FI93} and in \cite{GP98}, we deduce the insider martingale measure $\mathbb Q$ 
which plays an important role in the study of (semi)martingale processes in the filtration $\mathbb G$. We give the decomposition formula of an $\mathbb F$-martingale as a semimartingale in $\mathbb G$, which gives a positive answer to the Jacod's (H')-hypothesis and allows us to characterize the $\mathbb G$-portfolio wealth processes as in \cite{AIS98}.

In the optimization problem with random default times, it is often supposed that the random time satisfies the intensity hypothesis (e.g. Lim and Quenez \cite{LQ2011} and Kharroubi et al. \cite{KLN2013}) or the density hypothesis (e.g. Blanchet-Scalliet et al. \cite{BEJL2008}, Jeanblanc et al. \cite{JMPR2015} and Jiao et al. \cite{JKP2013}), so that it is a totally inaccessible stopping time in the market filtration. In particular, in \cite{JKP2013}, we consider marked random times where the random mark represents the loss at default and we suppose that the couple of default time and mark admits a conditional density. In this current paper, the random time $\tau$ we consider does not necessarily satisfy the intensity  nor the density hypothesis: it is a general stopping time in $\mathbb F$ and may also contain a predictable part. 
We obtain the optimal strategy and wealth for the two types of investors with a logarithmic utility function and deduce the additional gain due to the extra information. As a concrete case, we consider a hybrid default model similar as in Campi et al. \cite{CPS09} where the filtration $\mathbb F$ is generated by  a Brownian motion and a Poisson process, and the default time is the minimum of two random times: the first hitting time of a Brownian diffusion and the first jump time of the Poisson process and we compute the additional expected logarithmic utility wealth. 

The rest of the paper is organized as following. We model in Section \ref{section: setup} the filtration which represents the default time together with the random mark and we study its theoretical properties. Section \ref{sec-log-opt} focuses on the logarithmic utility optimization problem for the insider and compares the result with the case for ordinary investor. In Section \ref{sec:hybrid model}, we present the optimization results for an explicit hybrid default model. Section \ref{sec:conclusion} concludes the paper.

\section{Model framework}\label{section: setup}
In this section, we present our model setup. In particular, we study the enlarged filtration including the random mark which is a mixture of the initial and the progressive enlargements of filtrations. 
\subsection{The enlarged filtration and martingale processes}
Let $(\Omega,\mathcal A,\mathbb P)$ be a probability space  equipped with a filtration $\mathbb F=(\mathcal F_t)_{t\geq 0}$ which satisfies the usual conditions and $\tau$ be an $\mathbb F$-stopping time. 
Let $G$ be a random variable valued in a measurable space $(E,\mathcal E)$ and $\mathbb G=(\mathcal G_t)_{t\geq 0}$ be the smallest filtration containing $\mathbb F$ such that  $G$ is $\mathcal G_\tau$-measurable. By definition, one has 
\begin{equation}\label{equ:filtrationG}\forall\,t\in\mathbb R_+,\quad\mathcal G_t=\mathcal F_{t}\vee\sigma\Big(\big\{A\cap\{\tau\leq s \}\,|\,A\in\sigma(G),\; s\leq t\big\}\Big).\end{equation}
In particular, similar as in Jeulin \cite{Jeulin} (see also Callegaro et al. \cite{CMZ2013}), a stochastic process $Z$ is $\mathbb G$-adapted if and only if it can be written in the form 
\begin{equation}\label{equ:decompositionzt}Z_t=\indic_{\{\tau>t\}}Y_t+\indic_{\{\tau\leq t\}}Y_t(G), \quad t\geq 0\end{equation}
where $Y$ is an $\mathbb F$-adapted process and $Y(\cdot)$ is an $\mathbb F\otimes\mathcal E$-adapted process on $\Omega\times E$, where $\mathbb F\otimes\mathcal E$ denotes the filtration $(\mathcal F_t\otimes\mathcal E)_{t\geq 0}$. The following proposition characterizes the $\mathbb G$-predictable processes. The proof combines the techniques in those of \cite{Jeulin} Lemma 3.13 and 4.4 and is postponed in Appendix.

\begin{proposition}\label{Pro:predictableprocess}Let $\mathcal P(\mathbb F)$ be the predictable $\sigma$-algebra of the filtration $\mathbb F$.
A $\mathbb G$-adapted process $Z$ is $\mathbb G$-predictable if and only if it can be written in the form 
\begin{equation}\label{equ:Zpredictable}
Z_t=\indic_{\{\tau\geq t\}}Y_t+\indic_{\{\tau<t\}}Y_t(G),\quad t> 0,
\end{equation}
where $Y$ is an $\mathbb F$-predictable process and $Y(\cdot)$ is a $\mathcal P(\mathbb F)\otimes\mathcal E$-measurable function.
\end{proposition}

We study the  martingale processes in the filtrations $\mathbb F$ and $\mathbb G$. One basic martingale in $\mathbb F$ is related to the random time $\tau$. Let $D=(\indic_{\{\tau\leq t\}},\,t\geq 0)$ be the indicator process of the $\mathbb F$-stopping time $\tau$. Recall that the $\mathbb F$-compensator process $\Lambda$ of $\tau$ is the $\mathbb F$-predictable increasing process $\Lambda$ such that $N:=D-\Lambda$ is an $\mathbb F$-martingale. In particular, if $\tau$ is a predictable $\mathbb F$-stopping time, then $\Lambda$ coincides with $D$.

To study $\mathbb G$-martingales, we assume the following hypothesis for the random variable $G$ with respect to the filtration $\mathbb F$ (c.f. \cite{GP98} in the initial enlargement setting, see also \cite{Jacod} for comparison).
\begin{assumption}\label{Assum:Jacod} For any $t\geq 0$, the $\mathcal F_t$-conditional law of $G$ is equivalent to the probability law $\eta$ of $G$, i.e., $\mathbb P(G\in\cdot|\mathcal F_t)\sim\eta(\cdot)$, a.s.. Moreover, we 
denote by $p_t(\cdot)$ the conditional density  \begin{equation}\label{equ: jacod hypothesis}\mathbb P(G\in dx|\mathcal F_t)=p_t(x)\eta(dx),\quad a.s..\end{equation}
\end{assumption}

As pointed out in \cite[Lemma 1.8]{Jacod}, we can choose a version of the conditional probability density $p(\cdot)$, such that $p_t(\cdot)$ is $\mathcal F_t\otimes\mathcal E$-measurable for any $t\geq 0$ and that $(p_t(x),\,t\geq 0)$ is a positive c\`adl\`ag $(\mathbb F,\mathbb P)$-martingale for any $x\in E$. In the following we will fix such a version of the conditional density.  

\begin{remark}\rm
We assume the hypothesis of Jacod which is widely adopted in the study of initial and progressive enlargements of filtrations. Compared to the standard initial enlargement of $\mathbb F$ by $G$, the information of the random variable $G$ is added at a random time $\tau$ but not at the initial time; compared to the progressive enlargement, the random variable added here is the associated information $G$ instead of the  random time $\tau$. In particular, the behavior of $\mathbb G$-martingales is quite different from  the classic settings, and worth to be examined in detail.
\end{remark}


Similar as in \cite{FI93} and \cite{GP98}, we introduce the insider martingale measure $\mathbb Q$ which will be useful in the sequel. 
\begin{proposition}\label{pro:changementproba}
There exists a unique probability measure $\mathbb Q$ on $\mathcal F_\infty\vee\sigma(G)$ which verifies the following conditions:
\begin{enumerate}[(1)]
\item the probability measures $\mathbb Q$ and $\mathbb P$ are equivalent;
\item $\mathbb Q$ identifies with $\mathbb P$ on $\mathbb F$ and on $\sigma(G)$;
\item $G$ is independent of $\mathbb F$ under the probability $\mathbb Q$.
\end{enumerate}
Moreover, the Radon-Nikodym density of $\mathbb Q$ with respect to $\mathbb P$ on $\mathcal G_t$ is given by
\begin{equation}\label{equ:rnqp}\frac{d\mathbb Q}{d\mathbb P}\Big|_{\mathcal G_t}=\indic_{\{\tau>t\}}+\indic_{\{\tau\leq t\}}p_t(G)^{-1}\end{equation}
\end{proposition}
\bproof Let $\mathbb Q$ be defined by 
\[\frac{d\mathbb Q}{d\mathbb P}\Big|_{\mathcal F_t\vee\sigma(G)}={p_t(G)^{-1}},\quad t\geq 0.\]Since $(\mathcal F_t\vee\sigma(G))_{t\geq 0}$ is the initial enlargement of $\mathbb F$ by $G$,  we obtain by \cite{FI93, GP98} that $\mathbb Q$ is the unique equivalent probability measure on $\mathcal F_\infty\vee\sigma(G)$ which satisfies the conditions (1)--(3). Moreover, the Radon-Nikodym density $d\mathbb Q/d\mathbb P$ on $\mathcal G_t$ is given by
\[\mathbb E^{\mathbb P}[p_t(G)^{-1}|\mathcal G_t]=\mathbb E^{\mathbb P}[\indic_{\{\tau>t\}}p_t(G)^{-1}|\mathcal G_t]+\indic_{\{\tau\leq t\}}p_t(G)^{-1}.\]
Let $Z_t$ be a bounded $\mathcal G_t$-measurable random variable. By the decomposed form \eqref{equ:decompositionzt} we obtain that $\indic_{\{\tau>t\}}Z_t$ is $\mathcal F_t$-measurable. Hence
\[\mathbb E^{\mathbb P}[\indic_{\{\tau>t\}}p_t(G)^{-1}Z_t]=\mathbb E^{\mathbb P}[\indic_{\{\tau>t\}}Z_t\mathbb E^{\mathbb P}[p_t(G)^{-1}|\mathcal F_t]],\]
which leads to
\[\mathbb E^{\mathbb P}[\indic_{\{\tau>t\}}p_t(G)^{-1}|\mathcal G_t]=\indic_{\{\tau>t\}}\mathbb E^{\mathbb P}[p_t(G)^{-1}|\mathcal F_t]=\indic_{\{\tau>t\}}\quad\text{a.s.}\] 
Hence we obtain \eqref{equ:rnqp}.
\finproof

The following proposition shows that the filtration $\mathbb G$ also satisfies the usual conditions under the $\mathbb F$-density hypothesis on the random variable $G$. The idea follows \cite[Proposition 3.3]{Amen2000}. 
\begin{proposition}
Under Assumption \ref{Assum:Jacod}, the enlarged filtration $\mathbb G$ is right continuous.
\end{proposition}
\bproof
The statement does not involve the underlying probability measure. Hence we may assume without loss of generality (by Proposition \ref{pro:changementproba}) that $G$ is independent of $\mathbb F$ under the probability $\mathbb P$. Let $t\geqslant 0$ and $\varepsilon>0$. Let $X_{t+\varepsilon}$ be a bounded $\mathcal G_{t+\varepsilon}$-measurable random variable. We write it in the form
\[X_{t+\varepsilon}=Y_{t+\varepsilon}\indic_{\{\tau>t+\varepsilon\}}+Y_{t+\varepsilon}(G)\indic_{\{\tau\leq t+\varepsilon\}},\] 
where $Y_{t+\varepsilon}$ and $Y_{t+\varepsilon}(\cdot)$ are respectively bounded $\mathcal F_{t+\varepsilon}$-measurable and $\mathcal F_{t+\varepsilon}\otimes\mathcal E$-measurable functions. Then for $\delta\in(0,\varepsilon)$, by the independence between $G$ and $\mathbb F$ one has
\[\begin{split}\mathbb E^{\mathbb P}[X_{t+\varepsilon}\,|\,\mathcal G_{t+\delta}]&=\indic_{\{\tau>t+\delta\}}{\mathbb E^{\mathbb P}[Y_{t+\varepsilon}\indic_{\{\tau>t+\varepsilon\}}+Y_{t+\varepsilon}(G)\indic_{\{t+\delta<\tau\leq t+\varepsilon\}}\,|\,\mathcal F_{t+\delta}]}\\
&\qquad+\indic_{\{\tau\leq t+\delta\}}\mathbb E^{\mathbb P}[Y_{t+\varepsilon}(x)\,|\,\mathcal F_{t+\delta}]_{x=G}\\
&=\indic_{\{\tau>t+\delta\}}\Big(\mathbb E^{\mathbb P}[Y_{t+\varepsilon}\indic_{\{\tau>t+\varepsilon\}}\,|\,\mathcal F_{t+\delta}]+\int_{E}\mathbb E^{\mathbb P}[Y_{t+\varepsilon}(x)\indic_{\{t+\delta<\tau\leq t+\varepsilon\}}\,|\,\mathcal F_{t+\delta}]\,\eta(dx)\Big)\\
&\qquad+\indic_{\{\tau\leq t+\delta\}}\mathbb E^{\mathbb P}[Y_{t+\varepsilon}(x)\,|\,\mathcal F_{t+\delta}]_{x=G},
\end{split}\]
where $\eta$ is the probability law of $G$.
Since the filtration $\mathbb F$ satisfies the usual conditions, any $\mathbb F$-martingale admits a c\`adl\`ag version. Therefore, by taking a suitable version of the expectations $\mathbb E^{\mathbb P}[X_{t+\varepsilon}\,|\,\mathcal F_{t+\delta}]$, we have
\[\begin{split}\lim_{\delta\rightarrow 0}\mathbb E^{\mathbb P}[X_{t+\varepsilon}\,|\,\mathcal G_{t+\delta}]=&\indic_{\{\tau>t\}}\Big(\mathbb E^{\mathbb P}[Y_{t+\varepsilon}\indic_{\{\tau>t+\varepsilon\}}\,|\,\mathcal F_t]+\int_E\mathbb E^{\mathbb P}[Y_{t+\varepsilon}(x)\indic_{\{t<\tau\leq t+\varepsilon\}}\,|\,\mathcal F_t]\,\eta(dx)\Big)\\
&\qquad +\indic_{\{\tau\leq t\}}\mathbb E^{\mathbb P}[Y_{t+\varepsilon}(x)\,|\,\mathcal F_t]_{x=G}=\mathbb E^{\mathbb P}[X_{t+\varepsilon}\,|\,\mathcal G_t].\end{split}\]
In particular, if $X$ is a bounded $\mathcal G_{t+}:=\bigcap_{\varepsilon>0}\mathcal G_{t+\varepsilon}$-measurable random variable, then one has $\mathbb E^{\mathbb P}[X\,|\,\mathcal G_t]=X$ almost surely. Hence $\mathcal G_{t+}=\mathcal G_t$.
\finproof

Under the probability measure $\mathbb Q$, the random variable $G$ is independent of $\mathbb F$. This observation leads to the following characterization of $(\mathbb G,\mathbb Q)$-(local)-martingales. In the particular case where $\tau=0$, we recover the classic result on initial enlargement of filtrations.

\begin{proposition}\label{pro:criteremartingale}
Let $Z=(\indic_{\{\tau>t\}}Y_t+\indic_{\{\tau\leq t\}}Y_t(G), t\geq 0)$ be a $\mathbb G$-adapted process. We assume that \begin{enumerate}[(1)]
\item  $Y(\cdot)$ is an $\mathbb F\otimes\mathcal E$-adapted process such that $Y(x)$ is an $(\mathbb F,\mathbb P)$-square-integrable martingale for any $x\in E$ (resp. an $(\mathbb F,\mathbb P)$-locally square-integrable martingale with a common localizing stopping time sequence independent of $x$),
\item the process
\[\widetilde Y_t:=\indic_{\{\tau>t\}}Y_t+\int_E\Big(\int_{]0,t]}Y_{u-}(x)\,d\Lambda_u+\langle N,Y(x)\rangle_t^{\mathbb F,\mathbb P}\Big)\eta(dx), \quad t\geq 0\]
is well defined and is an $(\mathbb F,\mathbb P)$-martingale (resp. an $(\mathbb F,\mathbb P)$-local martingale).
\end{enumerate}
Then the process $Z$ is a $(\mathbb G,\mathbb Q)$-martingale (resp. a $(\mathbb G,\mathbb Q)$-local martingale).
\end{proposition}
\bproof
We can reduce the local martingale case to the martingale case by taking a sequence of $\mathbb F$-stopping times which localizes the processes appearing in the conditions (1) and (2). Therefore, we only treat the martingale case. 
Note that since $N$  and $Y(x)$ are square integrable (c.f. \cite[Chapitre VII (15.1)]{DMII} for the square integrability of $N$),  $NY(x)-\langle N,Y(x)\rangle^{\mathbb F,\mathbb P}$ is an $(\mathbb F,\mathbb P)$-martingale by \cite[Chapter I, Theorem 4.2]{JS03}.

\noindent For $t\geq s\geq 0$, one has
\begin{equation}\label{equ:ZQmartingale}\begin{split}&\quad\;\mathbb E^{\mathbb Q}[Z_t|\mathcal G_s]=\mathbb E^{\mathbb Q}[\indic_{\{\tau>t\}}Y_t|\mathcal G_s]+\mathbb E^{\mathbb Q}[\indic_{\{\tau\leq t\}}Y_t(G)|\mathcal G_s]\\
&=\indic_{\{\tau>s\}}\Big(\mathbb E^{\mathbb Q}[\indic_{\{\tau>t\}}Y_t|\mathcal F_s]+\int_E\mathbb E^{\mathbb Q}[\indic_{\{s<\tau\leq t\}}Y_t(x)|\mathcal F_s]\,\eta(dx)\Big)+\indic_{\{\tau\leq s\}}\mathbb E^{\mathbb Q}[Y_t(x)|\mathcal F_s]|_{x=G}\\
&=\indic_{\{\tau>s\}}\Big(\mathbb E^{\mathbb P}[\indic_{\{\tau>t\}}Y_t|\mathcal F_s]+\int_E\mathbb E^{\mathbb P}[\indic_{\{s<\tau\leq t\}}Y_t(x)|\mathcal F_s]\,\eta(dx)\Big)+\indic_{\{\tau\leq s\}}\mathbb E^{\mathbb P}[Y_t(x)|\mathcal F_s]|_{x=G}\\
&=\indic_{\{\tau>s\}}\Big(\mathbb E^{\mathbb P}[\indic_{\{\tau>t\}}Y_t|\mathcal F_s]+\int_E\mathbb E^{\mathbb P}[\indic_{\{s<\tau\leq t\}}Y_t(x)|\mathcal F_s]\,\eta(dx)\Big)+\indic_{\{\tau\leq s\}}Y_s(G)
\end{split}\end{equation}
where the second equality comes from the fact that $G$ is indenpendent of $\mathbb F$ under the probability $\mathbb Q$ and that $\eta$ coincides with the $\mathbb Q$-probability law of $\mathbb G$, and the third equality comes from the fact that the probability measures $\mathbb P$ and $\mathbb Q$ coincide on the filtration $\mathbb F$.

 Since $Y(x)$ is an $(\mathbb F,\mathbb P)$-martingale, one has
\begin{equation}\label{equ:espytfs}\begin{split}&\quad\;\mathbb E^{\mathbb P}[\indic_{\{s<\tau\leq t\}}Y_t(x)\,|\,\mathcal F_s]=\mathbb E^{\mathbb P}[D_tY_t(x)|\mathcal F_s]-D_sY_s(x)\\&=\mathbb E^{\mathbb P}[N_tY_t(x)-N_sY_s(x)|\mathcal F_s]+\mathbb E^{\mathbb P}[\Lambda_tY_t(x)-\Lambda_sY_s(x)|\mathcal F_s]\\
&=\mathbb E^{\mathbb P}[\langle N,Y(x)\rangle_t^{\mathbb F,\mathbb P}-\langle N,Y(x)\rangle_s^{\mathbb F,\mathbb P}|\mathcal F_s]+\mathbb E^{\mathbb P}[\Lambda_tY_t(x)-\Lambda_sY_s(x)|\mathcal F_s],
\end{split}\end{equation}
where the last equality comes from the fact that $NY(x)-\langle N,Y(x)\rangle^{\mathbb F,\mathbb P}$ is an $(\mathbb F,\mathbb P)$-martingale. Moreover, since $Y(x)$ is an $(\mathbb F,\mathbb P)$-martingale, its predictable projection is $Y_{-}(x)$ (see \cite[Chapter I, Corollary 2.31]{JS03}), and hence
\begin{equation}\label{equ:esplamdatyt}\mathbb E^{\mathbb P}[\Lambda_tY_t(x)-\Lambda_sY_s(x)\,|\,\mathcal F_s]=\mathbb E\bigg[\int_{(s,t]}Y_{u-}(x)\,d\Lambda_u\,\bigg|\,\mathcal F_s\bigg]\end{equation}
since $\Lambda$ is an integrable increasing process which is $\mathbb F$-predictable (see \cite[VI.61]{DMII}). Therefore, by \eqref{equ:ZQmartingale} we obtain
\[\begin{split}&\quad\;\mathbb E^{\mathbb Q}[Z_t|\mathcal G_s]-Z_s\\
&=\indic_{\{\tau>s\}}\bigg(\mathbb E^{\mathbb P}[\indic_{\{\tau>t\}}Y_t-\indic_{\{\tau>s\}}Y_s|\mathcal F_s]+\int_E\mathbb E^{\mathbb P}\bigg[\langle N,Y(x)\rangle^{\mathbb F,\mathbb P}_t-\langle N,Y(x)\rangle^{\mathbb F,\mathbb P}_s+\int_{(s,t]}Y_{u-}(x)d\Lambda_u\bigg|\mathcal F_s\bigg]\eta(dx)\bigg)\\
&=\indic_{\{\tau>s\}}\mathbb E^{\mathbb P}[\widetilde Y_t-\widetilde Y_s\,|\,\mathcal F_s]=0.
\end{split}\]
The proposition is thus proved.
\finproof

\begin{corollary}\label{cor:Gadaptedprocess}Let 
$Z=(\indic_{\{\tau>t\}}Y_t+\indic_{\{\tau\leq t\}}Y_t(G),\, t\geq 0)$
be a $\mathbb G$-adapted process.
Then $Z$ is a $(\mathbb G,\mathbb P)$-martingale (resp. local $(\mathbb G,\mathbb P)$-martingale) if the following conditions are fulfilled:
\begin{enumerate}[(1)]
\item for any $x\in E$, $(Y_t(x)p_t(x), t\geq 0)$ is an $(\mathbb F,\mathbb P)$-square integrable martingale (resp. a $(\mathbb F,\mathbb P)$-locally square integrable martingale with a common localizing stopping time sequence);
\item the process
\[\indic_{\{\tau>t\}}Y_t+\int_E\Big(\int_{]0,t]}Y_{u-}(x)p_{u-}(x)\,d\Lambda_u+\langle N,Y(x)p(x)\rangle_t^{\mathbb F,\mathbb P}\Big)\eta(dx), \quad t\geq 0\]
is a $(\mathbb F,\mathbb P)$-martingale (resp. a local $(\mathbb F,\mathbb P)$-martingale).
\end{enumerate}
\end{corollary}
\bproof
By Proposition \ref{pro:changementproba}, $Z$ is a $(\mathbb G,\mathbb P)$-(local)-martingale if and only if the process $Z(\indic_{[\![0,\tau[\![}+\indic_{[\![\tau,+\infty[\![}\,p(G))$ is a $(\mathbb G,\mathbb Q)$-(local)-martingale. Therefore the assertion results from Proposition \ref{pro:criteremartingale}.
\finproof

\begin{proposition}\label{pro:characterzationGmartinga}
Let $Z$ be a $(\mathbb G,\mathbb P)$-martingale on $[0,T]$ such that the process $\indic_{[\![\tau,+\infty[\![}Zp(G)$ is bounded. Then there exists an $\mathbb F$-adapted process $Y$ and an $\mathbb F\otimes\mathcal E$-adapted process $Y(\cdot)$ such that $Z_t=\indic_{\{\tau>t\}}Y_t+\indic_{\{\tau\leq t\}}Y_t(G)$ and that the following conditions are fulfilled:
\begin{enumerate}[(1)]
\item for any $x\in E$, $(Y_t(x)p_t(x), t\geq 0)$ is a bounded $(\mathbb F,\mathbb P)$-martingale;
\item the process
\[\indic_{\{\tau>t\}}Y_t+\int_E\Big(\int_{]0,t]}Y_{u-}(x)p_{u-}(x)\,d\Lambda_u+\langle N,Y(x)p(x)\rangle_t^{\mathbb F,\mathbb P}\Big)\eta(dx), \quad t\geq 0\]
is well defined and is an $(\mathbb F,\mathbb P)$-martingale.
\end{enumerate}
\end{proposition}
\bproof
Since $Z_T$ is a $\mathcal G_T$-measurable random variable, we can write it in the form 
\begin{equation}\label{equ:ZT}Z_T=\indic_{\{\tau>T\}}Y_T+\indic_{\{\tau\leq T\}}Y_T(G),\end{equation}
where $Y_T$ is an $\mathcal F_T$-measurable random variable, and $Y_T(\cdot)$ is an $\mathcal F_T\otimes\mathcal E$-measurable function such that $Y_T(\cdot)p_T(\cdot)$ is bounded. Similarly to \cite[Lemma 1.8]{Jacod}, we can construct an $\mathbb F\otimes\mathcal E$-adapted process $Y(\cdot)$ on $[0,T]$ such that  $Y(x)p(x)$ is a c\`adl\`ag $(\mathbb F,\mathbb P)$-martingale for any $x\in E$. In particular, for $t\in[0,T]$ one has
\begin{equation}\label{equ:ytx}Y_t(x)=\mathbb E^{\mathbb P}\bigg[\frac{Y_T(x)p_T(x)}{p_t(x)}\bigg|\mathcal F_t\bigg].\end{equation}
We then let, for $t\in[0,T]$ 
\begin{equation}\label{equ:ytildet}\widetilde Y_t:=\mathbb E^{\mathbb P}\bigg[Y_T\indic_{\{\tau>T\}}+\int_E\bigg(\int_{]0,T]}Y_{u-}(x)p_{u-}(x)\,d\Lambda_u+\langle N,Y(x)p(x)\rangle^{\mathbb F,\mathbb P}_T\bigg)\eta(dx)\bigg|\mathcal F_t\bigg].\end{equation}
Then $\widetilde Y$ is an $(\mathbb F,\mathbb P)$-martingale. For any $t\in [0,T]$, we let $Y_t$ be an $\mathcal F_t$-measurable random variable such that
\[\indic_{\{\tau>t\}}Y_t=\widetilde Y_t-\int_E\Big(\int_{]0,t]}Y_{u-}(x)p_{u-}(x)\,d\Lambda_u+\langle N,Y(x)p(x)\rangle^{\mathbb F,\mathbb P}_t\Big)\eta(dx).\]
This is always possible since
\[\begin{split}&\quad\;\indic_{\{\tau\leq t\}}\bigg(\widetilde Y_t-\int_E\Big(\int_{]0,t]}Y_{u-}(x)p_{u-}(x)\,d\Lambda_u+\langle N,Y(x)p(x)\rangle^{\mathbb F,\mathbb P}_t\Big)\eta(dx)\bigg)\\
&=\indic_{\{\tau\leq t\}}\mathbb E^{\mathbb P}\bigg[\int_E\Big(\int_{]t,T]}Y_{u-}(x)p_{u-}(x)\,d\Lambda_u+d\langle N,Y(x)p(x)\rangle^{\mathbb F,\mathbb P}_u\Big)\eta(dx)\bigg|\mathcal F_t\bigg]\\
&=\indic_{\{\tau\leq t\}}\int_E\mathbb E^{\mathbb P}[\indic_{\{t<\tau\leq T\}}Y_T(x)p_T(x)\,|\,\mathcal F_t]\,\eta(dx)=0,
\end{split}\] 
where the second equality is obtained by an argument similar to \eqref{equ:espytfs} and \eqref{equ:esplamdatyt}. 
We finally show that $Z_t=\indic_{\{\tau>t\}}Y_t+\indic_{\{\tau\leq t\}}Y_t(G)$ $\mathbb P$-a.s. for any $t\in[0,T]$. Note that we already have $Z_T=\indic_{\{\tau>t\}}Y_T+\indic_{\{\tau\leq T\}}Y_T(G)$. Therefore it remains to prove that the $\mathbb G$-adapted process $(\indic_{\{\tau>t\}}Y_t+\indic_{\{\tau\leq t\}}Y_t(G))_{t\in[0,T]}$ is an $(\mathbb G,\mathbb P)$-martingale. This follows  from the construction of the processes $Y$, $Y(\cdot)$ and Corollary \ref{cor:Gadaptedprocess}.
\finproof

\begin{remark}\rm
\begin{enumerate}[(1)]
\item We observe from the proof of the previous proposition that, if $Z$ is a $(\mathbb G,\mathbb P)$-martingale on $[0,T]$ (without boundedness hypothesis) such that $Z_T$ can be written into the form \eqref{equ:ZT} with $Y_T(x)p_T(x)\in L^2(\Omega,\mathcal F_T,\mathbb P)$ for any $x\in E$, then we can construct the $\mathbb F\otimes\mathcal E$-adapted process $Y(\cdot)$ by using the relation \eqref{equ:ytx}. Note that for any $x\in E$, the process $Y(x)p(x)$ is a square-integrable $(\mathbb F,\mathbb P)$-martingale. Therefore, the result of Proposition \ref{pro:characterzationGmartinga} remains true provided that the conditional expectation in \eqref{equ:ytildet} is well defined.
\item Let $Z$ be a $(\mathbb G,\mathbb P)$-martingale on $[0,T]$. In general, the decomposition of $Z$ into the form $Z=\indic_{[\![0,\tau[\![}Y+\indic_{[\![\tau,+\infty[\![}Y(G)$ with $Y$ being $\mathbb F$-adapted and $Y(\cdot)$ being $\mathbb F\otimes\mathcal E$-adapted is not unique. Namely, there may exist an $\mathbb F$-adapted process $\widetilde Y$ and an $\mathbb F\otimes\mathcal E$-adapted process $\widetilde Y(\cdot)$ such that $\widetilde Y$ is not a version of $Y$, $\widetilde Y(\cdot)$ is not a version of $Y(\cdot)$, but we still have $Z=\indic_{[\![0,\tau[\![}\widetilde Y+\indic_{[\![\tau,+\infty[\![}\widetilde Y(G)$. 
 Moreover, although the proof of Proposition \ref{pro:characterzationGmartinga} provides an explicit way to construct the decomposition of the $(\mathbb G,\mathbb P)$-martingale $Z$ which satisfies the two conditions,  in general such decomposition is not unique neither. 
\item Concerning the local martingale analogue of Proposition \ref{pro:characterzationGmartinga}, the main difficulty is that a local $(\mathbb G,\mathbb P)$-martingale need not be localized  by a sequence of $\mathbb F$-stopping times. To solve this problem, it is crucial to understand the $\mathbb G$-stopping times and their relation with $\mathbb F$-stopping times.
\end{enumerate}
\end{remark}

\subsection{(H')-hypothesis and semimartingale decomposition}

In this subsection, we prove that under Assumption \ref{Assum:Jacod}, the (H')-hypothesis (see \cite{Jacod}) is satisfied and we give the semimartingale decomposition  of an $\mathbb F$-martingale in $\mathbb G$. 

\begin{theorem}\label{thm-decomp}We suppose that Assumption \ref{Assum:Jacod} holds.
Let $M$ be an $(\mathbb F,\mathbb P)$-locally square integrable martingale, then it is a $(\mathbb G,\mathbb P)$-semimartingale. Moreover, the process
\[\widetilde M_t=M_t-\indic_{\{\tau\leq t\}}\int_{]0,t]}\frac{d\langle M-M^\tau,p(x)\rangle_s^{\mathbb F,\mathbb P}}{p_{s-}(x)}\bigg|_{x=G}\,\,, \quad t\geq 0\]
is a $(\mathbb G,\mathbb P)$-local martingale, where $M^\tau=(M_t^{\tau},t\geq 0)$ is the stopped process with $M_t^{\tau}=M_{t\wedge\tau}$.
\end{theorem}

We present two proofs of Theorem \ref{thm-decomp}. The first one relies on the following Lemma, which computes the $(\mathbb G,\mathbb Q)$-predictable bracket of an $(\mathbb F,\mathbb P)$-local martingale with a general $(\mathbb \F,\mathbb P)$-local martingale. This approach is more computational, but Lemma \ref{Lem:mainlemma} has its own interest, in particular for the study of $\mathbb G$-adapted processes. The second proof is more conceptual and relies on a classic result of Jacod \cite{Jacod} on initial enlargement of filtrations under an additional (positivity or integrability) assumption on the process $\widetilde M$.
\begin{lemma}\label{Lem:mainlemma}
Let $Y$ be an $\mathbb F$-adapted process and $Y(\cdot)$ be an $\mathbb F\otimes\mathcal E$-adapted process such that (as in Proposition \ref{pro:criteremartingale}) 
\begin{enumerate}[(1)]
\item $Y(x)$ is an $(\mathbb F,\mathbb P)$-locally square integrable martingale for any $x\in E$,
\item the process
\[H_t:=\int_E\Big(\int_{]0,t]}Y_{u-}(x)d\Lambda_u+\langle N,Y(x)\rangle_t^{\mathbb F,\mathbb P}\Big)\,\eta(dx),\quad t\geq 0\]
is well defined and of finite variation, and 
$\widetilde Y=\indic_{[\![0,\tau[\![}Y+H$ is an  $(\mathbb F,\mathbb P)$-locally square-integrable martingale.
\end{enumerate}
Let $Z$ be the process $\indic_{[\![0,\tau[\![}Y+\indic_{[\![\tau,+\infty[\![}Y(G)$. Then one has
\begin{equation}\label{Equ:crochet}\begin{split}\langle M,Z\rangle_t^{\mathbb G,\mathbb Q}&=\langle M^\tau,\widetilde Y\rangle^{\mathbb F,\mathbb P}_t-\int_{]0,t]}M^\tau_{s-}\,d H_s+\int_E\Big(\int_{]0,t]}U_{s-}(x)d\Lambda_s+\langle N,U(x)\rangle_t^{\mathbb F,\mathbb P}\bigg)\eta(dx)\\
&\qquad +\langle M-M^\tau,Y(x)\rangle^{\mathbb F,\mathbb P}\Big|_{x=G},
\end{split}\end{equation}
where 
\[U_t(x)=M^\tau_t Y_t(x)-\langle M^\tau,Y(x)\rangle^{\mathbb F,\mathbb P}_t+\mathbb E^{\mathbb P}[\indic_{\{\tau<+\infty\}}\langle M^\tau,Y(x)\rangle_\tau^{\mathbb F,\mathbb P}|\mathcal F_t],\;x\in E.\]
\end{lemma}
\bproof 
It follows from Proposition \ref{pro:criteremartingale} that $Z$ is a $(\mathbb G,\mathbb Q)$-martingale. In the following, we establish the equality \eqref{Equ:crochet}. 

We first treat the case where the martingale $M$ begins at $\tau$ with $M_\tau=0$, namely  $M_t\indic_{\{\tau\geq t\}}=0$ for any $t\geq 0$. Therefore $W(x):=MY(x)-\langle M,Y(x)\rangle^{\mathbb F,\mathbb P}$ is a local $(\mathbb F,\mathbb P)$-martingale which vanishes on $[\![0,\tau]\!]$. In particular one has
\[\int_{]0,t]}W_{u-}(x)\,d\Lambda_u=0\quad\text{and}\quad\langle N,W(x)\rangle^{\mathbb P,\mathbb F}=0\]
since both processes $N$ and $\Lambda$ are stopped at $\tau$. By Proposition \ref{pro:criteremartingale}, we obtain that the process $W(G)=\indic_{[\![\tau,+\infty[\![}W(G)$ is actually a local $(\mathbb G,\mathbb Q)$-martingale. Note that 
\[W(G)= MY(G)-\langle M,Y(x)\rangle^{\mathbb F,\mathbb P}\big|_{x=G},\]
and $\langle M,Y(x)\rangle^{\mathbb P,\mathbb F}\big|_{x=G}$ is $\mathbb G$-predictable (by Proposition \ref{Pro:predictableprocess}, we also use the fact that $\langle M,Y(x)\rangle^{\mathbb P,\mathbb F}$ vanishes on $[\![0,\tau]\!]$), therefore we obtain $\langle M,Z\rangle^{\mathbb G,\mathbb Q}=\langle M,Y(x)\rangle^{\mathbb P,\mathbb F}\big|_{x=G}$.

In the second step, we assume that $M$ is stopped at $\tau$. In this case one has \[\forall\,t\geq 0,\quad U_t(x)=M_tY_t(x)-\langle M,Y(x)\rangle^{\mathbb F,\mathbb P}_t+\mathbb E^{\mathbb P}[\indic_{\{\tau<+\infty\}}\langle M,Y(x)\rangle_\tau^{\mathbb P,\mathbb F}|\mathcal F_t].\] 
It is a local $(\mathbb F,\mathbb P)$-martingale. Moreover, since $M$ is stopped at $\tau$, also is $\langle M,Y(x)\rangle^{\mathbb F,\mathbb P}$. In particular, since $\indic_{\{\tau\leq t\}}\langle M,Y(x)\rangle_{\tau}^{\mathbb F,\mathbb P}$ is $\mathcal F_t$-measurable, one has
\[\forall\,t\geq 0,\quad \indic_{\{\tau\leq t\}}M_tY_t(G)=\indic_{\{\tau\leq t\}}U_t(G).\]
In addition, by definition $\widetilde Y=\indic_{[\![0,\tau[\![}Y+H$. Hence one has
\[M\indic_{[\![0,\tau[\![}Y=M(\widetilde Y-H)=(M\widetilde Y-\langle M,\widetilde Y\rangle^{\mathbb F,\mathbb P})+\langle M,\widetilde Y\rangle^{\mathbb F,\mathbb P}-M_-\cdot H-H_-\cdot M-[M,H],\]
where $M_-\cdot H$ and $H_-\cdot M$ denote respectively the integral processes
\[\int_0^tM_{s-}\,dH_s,\quad\text{and}\quad\int_0^tH_{s-} \,dM_s,\quad t\geq 0.\]
Since $H$ is a predictable process of finite variation and $M$ is an $\mathbb F$-martingale, the process $[M,H]$ is a local $\mathbb F$-martingale (see \cite{JS03} Chapter I, Proposition 4.49). In particular, 
\[M\indic_{[\![0,\tau[\![}Y-\langle M,\widetilde Y\rangle^{\mathbb F,\mathbb P}+M_-\cdot H\]
is a local $\mathbb F$-martingale. Let
\[A_t=\langle M,\widetilde Y\rangle^{\mathbb F,\mathbb P}_t-\int_{]0,t]}M_{s-}dH_s+\int_E\bigg(\int_{]0,t]}U_{s-}(x)\,d\Lambda_s+\langle N,U(x)\rangle_t^{\mathbb F,\mathbb P}\bigg)\eta(dx),\quad t\geq 0.\]
This is an $\mathbb F$-predictable process, and hence is $\mathbb G$-predictable. Moreover, this process is stopped at $\tau$. Let $V$ be the $(\mathbb F,\mathbb P)$-martingale defined as \begin{equation}\label{equ:Vtdef}V_t=\mathbb E^{\mathbb P}[A_\tau\indic_{\{\tau<+\infty\}}\,|\,\mathcal F_t],\quad t\geq 0.\end{equation} Note that $V_t\indic_{\{\tau\leq t\}}=A_\tau\indic_{\{\tau\leq t\}}=A_t\indic_{\{\tau\leq t\}}$. Hence
\[AD=VD=V_-\cdot D+D_-\cdot V+[V,D]=V_-\cdot N+V_-\cdot\Lambda+D_-\cdot V+[V,N]+[V,\Lambda],\]
where $D=(\indic_{\{\tau\leq t\}},\;t\geq 0)=N+\Lambda$.
In particular,
\[AD-V_-\cdot\Lambda-\langle V,N\rangle^{\mathbb F,\mathbb P}=V_-\cdot N+D_-\cdot V+([V,N]-\langle V,N\rangle^{\mathbb P,\mathbb F})+[V,\Lambda]\]
is a local  $(\mathbb F,\mathbb P)$-martingale.
Therefore, one has
\[\begin{split}&\quad
\;\indic_{\{\tau>t\}}(M_tY_t-A_t)+\int_E\bigg(\int_{]0,t]}(U_{s-}(x)-V_{s-})d\Lambda_s+\langle N,U(x)-V\rangle_t^{\mathbb F,\mathbb P}\Bigg)\eta(dx)
\\
&=\indic_{\{\tau>t\}}M_tY_t-A_t+\indic_{\{\tau\leq t\}}A_t+\int_E\bigg(\int_{]0,t]}U_{s-}(x)\,d\Lambda_s+\langle N,U(x)\rangle_t^{\mathbb F,\mathbb P}\bigg)\eta(dx)-(V_-\cdot\Lambda)_t-\langle V,N\rangle_t^{\mathbb F,\mathbb P}\\
&=\Big(\indic_{\{\tau>t\}}M_tY_t-\langle M,\widetilde Y\rangle_t^{\mathbb F,\mathbb P}+(M_-\cdot H)_t\Big)+\Big(\indic_{\{\tau\leq t\}}A_t-(V_-\cdot\Lambda)_t-\langle V,N\rangle_t^{\mathbb F,\mathbb P} \Big),
\end{split}
\]
which is a local $(\mathbb F,\mathbb P)$-martingale. 

We write the process $MZ-A$ in the form
\[M_tZ_t-A_t=\indic_{\{\tau>t\}}(Y_tM_t-A_t)+\indic_{\{\tau\leq t\}}(U_t(G)-A_t)=\indic_{\{\tau>t\}}(Y_tM_t-A_t)+\indic_{\{\tau\leq t\}}(U_t(G)-V_t)\]
where the last equality comes from \eqref{equ:Vtdef}. We have seen that $U(x)-V$ is a local $(\mathbb F,\mathbb P)$-martingale for any $x\in E$. Hence by Proposition \ref{pro:criteremartingale} we obtain that $MZ-A$ is a local $(\mathbb G,\mathbb Q)$-martingale.

In the final step, we consider the general case. We decompose the $(\mathbb F,\mathbb P)$-martingale into the sum of two parts $M^\tau$ and $M-M^\tau$, where $M^\tau$ is an $(\mathbb F,\mathbb P)$-martingale stopped at $\tau$, and $M-M^\tau$ is an $(\mathbb F,\mathbb P)$-martingale which vanishes on $[\![0,\tau]\!]$. Combining the results obtained in the two previous steps, we obtain the formula \eqref{Equ:crochet}.
\finproof

\ni {\bf Proof of Theorem \ref{thm-decomp}.}
Since $\mathbb P$ and $\mathbb Q$ coincide on $\mathbb F$, we obtain that $M$ is an $(\mathbb F,\mathbb Q)$-martingale. Moreover, since $G$ is independent of $\mathbb F$ under the probability $\mathbb Q$, $M$ is also a $(\mathbb G,\mathbb Q)$-martingale.

We keep the notation of the Lemma \ref{Lem:mainlemma} and specify the terms in the situation of the theorem. Note that the Radon-Nikodym derivative of $\mathbb P$ with respect to $\mathbb Q$ on $\mathcal G_t$ equals 
\[Z_t:=\indic_{\{\tau>t\}}+\indic_{\{\tau\leq t\}}p_t(G).\]
In particular, with the notation of the lemma, one has
\[Y_t=1,\quad Y_t(x)=p_t(x),\quad t\geq 0,\;x\in E.\]
Since $\int_EY_t(x)\,\eta(dx)=1$, for any $t\geq 0$
\[\widetilde Y_t=\indic_{\{\tau>t\}}+\Lambda_t=1-N_t,\]
and
\[H_t=\int_E\bigg(\int_{]0,t]}Y_{u-}(x)d\Lambda_u+\langle N,Y(x)\rangle_t^{\mathbb F,\mathbb P}\bigg)\eta(dx)=\Lambda_t.\]
Moreover, one has
\[\int_EU_t(x)\eta(dx)=\int_E\Big(M^\tau_t Y_t(x)-\langle M^\tau,Y(x)\rangle_t^{\mathbb F,\mathbb P}+\mathbb E^{\mathbb P}[\langle M^\tau,Y(x)\rangle_\tau^{\mathbb F,\mathbb P}\indic_{\{\tau<+\infty\}}|\mathcal F_t]\Big)\eta(dx)=M^\tau.\]
Therefore, by Lemma \ref{Lem:mainlemma} one has
\[\begin{split}\langle M,Z\rangle^{\mathbb G,\mathbb Q}&=-\langle M^\tau,N\rangle^{\mathbb F,\mathbb P}-M_-^{\tau}\cdot\Lambda+M_-^{\tau}\cdot\Lambda+\langle M^\tau,N\rangle^{\mathbb F,\mathbb P}+\langle M-M^\tau,Y(x)\rangle^{\mathbb F,\mathbb P}\Big|_{x=G}\\
&=\langle M-M^\tau,Y(x)\rangle^{\mathbb F,\mathbb P}\Big|_{x=G}.
\end{split}\]
Finally, since $M$ is a $(\mathbb G,\mathbb Q)$-local martingale, by Girsanov's theorem (cf. \cite{JS03} Chapter III, Theorem 3.11), the process
\[\widetilde M_t=M_t-\int_{]0,t]}\frac{1}{Z_{s-}}d\langle M,Z\rangle^{\mathbb G,\mathbb Q}_s,\quad t\geq 0\]
is a local $(\mathbb G,\mathbb P)$-martingale. The theorem is thus proved.
\finproof

\ni {\bf Second proof of Theorem \ref{thm-decomp}.} Let $\mathbb H=\mathbb F\vee\sigma(G)$ be the initial enlargement of the filtration $\mathbb F$ by $\sigma(G)$. Clearly the filtration $\mathbb H$ is larger than $\mathbb G$. More precisely, the filtration $\mathbb G$ coincides with $\mathbb F$ before the stopping time $\tau$, and coincides with $\mathbb H$ after the stopping time $\tau$. We first observe that the stopped process at $\tau$ of an $(\mathbb F,\mathbb P)$-martingale $L$ is a $(\mathbb G,\mathbb P)$-martingale. In fact,  for $t\geq s\geq 0$ one has
\[\mathbb E[L^\tau_t|\mathcal G_s]=\indic_{\{\tau>s\}}\mathbb E[L_{\tau\wedge t}|\mathcal F_s]+\indic_{\{\tau\leq s\}}\mathbb E[L_\tau\,|\,\mathcal G_s]=\indic_{\{\tau>s\}}L_{s}+\indic_{\{\tau\leq s\}}L_\tau=L_{\tau\wedge s}.\]
We remark that, as shown by Jeulin's formula, this result holds more generally for any enlargement $\mathbb G$ which coincides with $\mathbb F$ before a random time $\tau$.

We now consider the decomposition of $M$ as $M=M^\tau+(M-M^\tau)$, where $M^{\tau}$ is the stopped process of $M$ at $\tau$. Since $\mathbb G$ coincides with $\mathbb F$ before $\tau$, we obtain by the above argument that $M^\tau$ is an $(\mathbb G,\mathbb P)$-local martingale. 
Consider now the process $Y:=M-M^\tau$, which begins at $\tau$. It is also an $(\mathbb F,\mathbb P)$-local martingale. By Jacod's decomposition formula (see \cite[Theorem 2.1]{Jacod}), the process
\begin{equation}\label{equ: Y compesated}\widetilde Y_t=Y_t-\int_{]0,t]}\frac{d\langle Y,p(x)\rangle_s^{\mathbb F,\mathbb P}}{p_{s-}(x)}\Big|_{x=G},\quad t\geq 0\end{equation}
is an $(\mathbb H,\mathbb P)$-local martingale. Note that the predictable quadratic variation process $\langle Y,p(x)\rangle_s^{\mathbb F,\mathbb P}$ vanishes on $[\![ 0,\tau]\!]$ since the process $Y$ begins at $\tau$. Hence
\[\int_{]0,t]}\frac{d\langle Y,p(x)\rangle_s^{\mathbb F,\mathbb P}}{p_{s-}(x)}=\indic_{\{\tau\leq t\}}\int_{]0,t]}\frac{d\langle Y,p(x)\rangle_s^{\mathbb F,\mathbb P}}{p_{s-}(x)}.\]
This observation also shows that the process \eqref{equ: Y compesated} is $\mathbb G$-adapted. Hence it is a $(\mathbb G,\mathbb P)$-local martingale under the supplementary assumption that $\widetilde Y$ is positive or $\|\widetilde Y\|_1<+\infty$, by Stricker \cite[Theorem 1.2]{stricker77}, where $\|\widetilde Y\|_1$ is defined as the supremum of $\|\widetilde Y_\sigma\|_{L^1}$ with $\sigma$ running over all finite $\mathbb G$-stopping times. Note that the condition $\|\widetilde Y\|_1<+\infty$ is satisfied if and only if the process $\widetilde Y$ is a $(\mathbb G,\mathbb P)$-quasimartingale (see \cite{Kazamaki72}).

\begin{remark}
\textcolor{red}{Even if if the  second proof of Theorem \ref{thm-decomp} needs the additional assumption on the positivity or integrability of the process $\widetilde M-M^\tau$, it remains interesting since it allows to weaken Assumption \ref{Assum:Jacod}. Indeed, to apply Jacod's decomposition formula we only need to assume that
the conditional law $\P(G\in.|\Fc_t)$ is  absolutely continuous w.r.t. $\P(G\in.)$.}
\end{remark}

\section{Logarithmic utility maximization}\label{sec-log-opt}

In this section,
we study the optimization problem for two types of investors: an insider and an ordinary agent. We consider a financial market composed by $d$ stocks with discounted prices given by the $d$-dimensional process $X=(X^1,\ldots,X^d)^\top$. This process is observed by both agents and is $\mathbb F$-adapted. We suppose that each $X^i$, $i=1,\ldots,d$, evolve according to the following stochastic differential equations
\beqs
 X^i_t & = & X_0^i+\int_0^tX^i_{s-} \Big( dM^i_s+\sum_{j=1}^d\alpha^j_s d\langle M^i, M^j\rangle_s\Big)\;,\quad t\geq 0\;,
\enqs
with $X^i_0$ a positive constant, $M^i$ an $\mathbb F$-locally square integrable martingale and $\alpha$ a $\Pc(\F)$--measurable process valued in $\R^d$ such that
\beq\label{cond-int-alpha}
\E\Big[\int_0^T\alpha^\top_s d\langle M\rangle_s \alpha_s\Big] & < & +\infty\;.
\enq
 The ordinary agent has access to the information flow given by the filtration $\F$, while the information flow of the insider is represented by the filtration $\G$. The optimization  for the ordinary agent is standard. For the insider, we follow \cite{AIS98, ABS2003} to solve the problem. We first describe the insider's portfolio in the enlarged filtration $\mathbb G$.
Recall that under Assumption \ref{Assum:Jacod}, the process $M$ is a $\G$-semimartingale with canonical decomposition given by Theorem \ref{thm-decomp}:
\beq\label{decompM}
M_t & = & \widetilde M _t +\indic_{\{\tau\leq t\}}\int_0^t\left. \frac{d\langle M-M^\tau,p(x)\rangle_s}{p_{s-}(x)}\right|_{x=G}\;,\quad t\geq 0\;,
\enq
where $\widetilde M$ is a $\G$-local martingale and $M^\tau$ is the stopped process $(M_{t\wedge\tau})_{t\geq 0}$.

\vspace{2mm}

Applying Theorem 2.5 of \cite{Jacod} to the $\F$-locally square integrable martingale $M-M^\tau$, we have the following result.

\begin{lemma}\label{Lem mi} For $i=1,\ldots,d$, there exists a $\Pc(\F)\otimes\Ec$-measurable function $m^i$ such that 
\beqs
\langle p(x),M^i-(M^i)^\tau\rangle_t & = & \int_0^tm^i_s(x)p_{s-}(x)d\langle M^i-(M^i)^\tau\rangle_s
\enqs
for all $x\in E$ and all $t\geq 0$.
\end{lemma}
We now rewrite the integral of $m$ w.r.t. $\langle M-M^\tau\rangle$.
\begin{lemma}\label{lem mu}
Under Assumption 2.1, there exists a $\Pc(\F)\otimes \Ec$-measurable process $\mu$ valued in $\R^d$ such that
\begin{equation*}
\int_0^t d\langle M-M^\tau\rangle_s \mu_s (x) =  
\left(
\begin{array}{c}
\int_0^t m_s^1(x)d\langle M^1-(M^1)^\tau\rangle_s
\\
\vdots\\
\int_0^t m_s^d(x)d\langle M^d-(M^d)^\tau\rangle_s
\end{array}
\right)
\end{equation*}
for all $t\geq 0$.
\end{lemma}
\ni \textbf{Proof.} The proof is the same as that of Lemma 2.8 in \cite{AIS98}. We therefore omit it.\finproof

\vspace{2mm}

We can then rewrite the process $M$ in \reff{decompM} in the following way
\beq\label{decompM with mu}
M_t & = & \widetilde M _t +\indic_{\{\tau\leq t\}}\int_0^t d\langle M-M^\tau\rangle_s\mu_{s}(G)\;,\quad t\geq 0\;,
\enq
and the dynamics of the process $X$ can be expressed with the $\G$-local martingale $\widetilde M$ as follows
\beqs
d X_t & = & \textrm{Diag}(X_{t-}) \Big( d\widetilde M_t+ d\langle M\rangle_t\alpha_t+d\langle M-M^\tau\rangle_t\mu_t(G)\Big)\;,\quad t\geq 0\;,
\enqs
where $\textrm{Diag}(X_{t-})$ stands for the $d\times d$ diagonal matrix  whose $i$-th diagonal term is $X^i_{t-}$ for $i=1,\ldots,d$.
We then introduce the following integrability assumption.
\begin{assumption}\label{Hyp-mu}
The process $\mu(G)$ is square integrable w.r.t. $d\langle M-M^\tau\rangle$:
\beqs
\E\Big[\int_0^T \mu_t(G)^\top d\langle M-M^\tau\rangle_t \mu_t(G)\Big] & < & \infty\;.
\enqs
\end{assumption}


\ni Denote by $\H\in\{\F,\G\}$ the underlying filtration. We define an $\H$-portfolio  as a couple $(x,\pi)$ where $x$ is a constant representing the initial wealth and $\pi$ is  an $\R^d$-valued $\Pc(\H)$-measurable process $\pi$ such that 
\beqs
\int_0^T\pi_t^\top d\langle M\rangle_t\pi_t & < & \infty \;,\quad \P\text{-a.s.}
\enqs
and
\begin{equation}\label{equ: investment strategy}\sum_{i=1}^d\pi^i_t\frac{\Delta X_t^i}{X_{t-}^i}>-1\;,\quad t\in[0,T]\;.\end{equation} 

Here $\pi^i_t$ represents the proportion of discounted wealth invested at time $t$ in the asset $X^i$.
For such an $\H$-portfolio, we define the associated discounted wealth process $V(x,\pi)$ by
\beqs
V_t(x,\pi) & = & x+\sum_{i=1}^d\int_0^t \pi_s^i \textcolor{red}{V_{s-}(x,\pi)} {d X^i_s\over X^i_{s-}}\;,\quad t\geq 0\;.
\enqs
By the condition \eqref{equ: investment strategy}, the wealth process is positive.
%
%
We suppose that the agents preferences are described by the logarithmic utility function. 
For a given initial capital $x$, we define the set of admissible $\H$-portfolio processes by
\beqs
\Ac^{\log}_\H(x) & = & \Big\{ \pi~:~ (x,\pi) \mbox{ is an $\H$-portfolio satisfying }\E\big[ \log ^- V_T(x,\pi)\big] <\infty\Big\}
\enqs
For an initial capital $x$ we then consider the two optimization problems:
\begin{itemize}
\item the ordinary agent's problem consists in computing
\beqs
V^{\log}_{\F} & = & \sup_{\pi\in\Ac^{\log}_\F(x)}\E\big[ \log V_T(x,\pi)\big]\;,
\enqs
\item the insider's problem consists in computing
\beqs
V^{\log}_{\G} & = & \sup_{\pi\in\Ac^{\log}_\G(x)}\E\big[ \log V_T(x,\pi)\big]\;.
\enqs
\end{itemize}
To solve these problems, we introduce the minimal martingale density processes $\hat Z^\F$ and  $\hat Z^\G$  defined by
\beqs
\hat Z ^\F_t & = & \mathscr{E} \Big(-\int_0^\cdot \alpha^\top_sdM_s\Big)_t
\enqs
and 
\beqs
\hat Z ^\G_t & = & \mathscr{E} \Big(-\int_0^\cdot \big(\alpha_s+\indic_{\tau\leq s}\mu_s(G)\big)^\top_sd\widetilde M_s\Big)_t
\enqs
for $t\in[0,T]$, where $\mathscr{E}(\cdot)$ denotes the Dol\' eans-Dade exponential. 
We first have the following result.
\begin{proposition}
(i) The processes $\hat Z ^\F X$ and $\hat Z ^\F V(x,\pi)$ are $\F$-local martingales for any portfolio $(x,\pi)$ such that $\pi\in\Ac_\F(x)$. 

\ni (ii) The processes $\hat Z ^\G X$ and $\hat Z ^\G V(x,\pi)$ are $\G$-local martingales for any portfolio $(x,\pi)$ such that $\pi\in\Ac_\G(x)$. 
\end{proposition}
\ni \textbf{Proof.} We only prove assertion (ii). The same arguments can be applied to prove (i) by taking $\mu(G)\equiv 0$. From It\-o's formula we have
\beqs
d(\hat Z^G X) & = & X_{-}d\hat Z^\G+\hat Z_{-}^\G  dX +d \langle Z^\G, X\rangle + d \big([ Z^\G, X]- \langle Z^\G, X\rangle\big)\;.
\enqs
From the dynamics of $\hat Z ^\G$ and $X$ we have
\beqs
d \langle \hat Z_{-}^\G, X\rangle & = & -\hat Z_{-}^\G\textrm{Diag}(X_{-})d\big\langle   \int_0^\cdot \big(\alpha_s+\indic_{\tau\leq s}\mu_s(G)\big)^\top_sd\widetilde M_s,M\big\rangle\\
 & = & -\hat Z_{-}^\G\textrm{Diag}(X_{-}) d\langle M \rangle \big(\alpha+\indic_{[\![\tau,+\infty[\![}\mu(G)\big)\\
  & = & -\hat Z_{-}^\G\textrm{Diag}(X_{-}) \big(d\langle M \rangle \alpha+d\langle M-M^\tau \rangle \mu(G)\big)\;.
\enqs
Therefore we get
\beqs
 d  (\hat Z^\G X)& = &  X_{-}d\hat Z^\G+\hat Z^\G_{-} \textrm{Diag}(X_{-}) d \widetilde M +d \big([ \hat Z^\G, X]- \langle \hat Z^\G, X\rangle \big)
\enqs
which shows that $\hat Z^\G X$ is a $\G$-local martingale. 
\ep

\vspace{2mm}

\ni We are now able to compute $V_\F$ and $V_\G$ and provide optimal strategies.

\begin{theorem}\label{THM ORD INS}
(i) An optimal strategy for the ordinary agent is given by
\beqs
\pi^{ord}_t  & = & \alpha_t \;,\quad t\in[0,T]\;, 
\enqs
and the maximal expected logarithmic utility is given by
\beqs
V^{\log}_\F & = & \E\Big[\log V_T\big(x,\pi^{ord}\big)\Big]~~=~~\log x +{1\over 2}\E\Big[\int_0^T\alpha_t^\top d \langle M\rangle _t\alpha_t\Big]\;.
\enqs
 (ii) An optimal strategy for the insider is given by
\beqs
\pi^{ins}_t  & = & \alpha_t +\indic_{\tau\leq t}\mu_t(G)\;,\quad t\in[0,T]\;, 
\enqs
and the maximal expected logarithmic utility is given by
 \beqs
V^{\log}_\G & = & \E\Big[\log V_T\big(x,\pi^{ins}\big)\Big]\\
 & = &\log x +{1\over 2}\E\Big[\int_0^T\alpha_t^\top d \langle M\rangle _t\alpha_t\Big]+{1\over 2}\E\Big[\int_0^T\mu_t(G)^\top d \langle M-M^\tau\rangle _t\mu_t(G)\Big]\;. 
 \enqs
 (iii) The insider's additional expected utility is given by
 \beqs
 V^{\log}_\G-V^{\log}_\F & = & {1\over 2}\E\Big[\int_0^T\mu_t(G)^\top d \langle M-M^\tau\rangle _t\mu_t(G)\Big]\;.
 \enqs
\end{theorem}
\ni \textbf{Proof.} We do not prove (i) since it relies on the same arguments as for (ii) with $\mu(G)\equiv 0$ and $\hat Z^\F$ in place of $\hat Z^\G$.  \ep

\ni (ii) We recall that for a $C^1$ concave function $u$ such that its derivative $u'$ admits an inverse function $I$ we have
\beqs
u(a) & \leq & u\big(I(b)\big)-b\big( I(b)-a \big)
\enqs
for all $a,b\in\R$. Applying this inequality with $u=\log$, $a=V_T(x,\pi)$ for $\pi\in\Ac_\G(x)$ and $b=y\hat Z_T^\G$ for some constant $y>0$ we get
\beqs
\log V_T(x,\pi) & \leq & \log {1\over y\hat Z_T^\G}- y\hat Z_T^\G\left({1\over y\hat Z_T^\G}-V_T(x,\pi)\right)\\
 & \leq & -\log y -\log \hat Z_T^\G-1 + y\hat Z_T^\G V_T(x,\pi)
\enqs
Since $V(x,\pi)$ is a non-negative process and $\hat Z^\G V(x,\pi)$ is a $\G$-local martingale it is a  $\G$-super-martingale. therefore, we get 
\beqs
\E\log V_T(x,\pi) & \leq & -1-\log y -\E\log \hat Z_T^\G +xy\;.
\enqs
Since this inequality holds for any $\pi\in\Ac_\G(x)$, we obtain by taking $y={1\over x}$
\beqs
V^{\log}_\G & \leq & \log x -\E\log \hat Z_T^\G\;.
\enqs
Moreover, we have
\beqs
\log V_T(x,\pi^{ins}) & = & \log x + \int_0^T {\pi_t^{ins}}^\top d\widetilde M_t+ \int_0^T{\pi_t^{ins}}^\top d \langle M\rangle_t \alpha_t+ \int_0^T{\pi_t^{ins}}^\top d \langle M-M^\tau\rangle_t \mu_t(G)\\
 & = & \log x + \int_0^T {\pi_t^{ins}}^\top d\widetilde M_t+ \int_0^T{\pi_t^{ins}}^\top d \langle M\rangle_t \alpha_t+ \int_0^T{\pi_t^{ins}}^\top d \langle M\rangle_t \mu_t(G)\indic_{\tau\leq t}\\
  & = & \log x + \int_0^T \big(\alpha_t+\mu_t(G)\big)^\top d\widetilde M_t\\
   & & +\int_0^T \big(\alpha_t+\mu_t(G)\indic_{\tau\leq t}\big)^\top d \langle M\rangle_t\big(\alpha_t+\mu_t(G)\indic_{\tau\leq t}\big)\\
   & = & \log x +\log \hat Z^\G_T
\enqs
From \reff{cond-int-alpha} and Assumption  \ref{Hyp-mu}, we get $\pi^{ins}\in \Ac^{\log}_\G(x)$. Therefore $\pi^{ins}$ is an optimal strategy for the insider's problem.

Using \reff{cond-int-alpha}, we get that $\int_0^. \alpha^\top d M$ and $\int_0^. \alpha^\top d \widetilde M$ are respectively $\F$ and $\G$ martingales. Therefore we have
\beqs
0~~=~~\E\Big[\int_0^T\alpha^\top d\widetilde M\Big]-\E\Big[\int_0^T\alpha^\top dM\Big] & = & \E\Big[\int_0^T\alpha^\top d\langle M\rangle \mu(G)\indic_{[\tau,+\infty)}\Big]\;,
\enqs
which gives
 \beqs
\E\Big[\log V_T\big(x,\pi^{ins}\big)\Big]
 & = &\log x +{1\over 2}\E\Big[\int_0^T\alpha_t^\top d \langle M\rangle _t\alpha_t\Big]+{1\over 2}\E\Big[\int_0^T\mu_t(G)^\top d \langle M-M^\tau\rangle _t\mu_t(G)\Big]\;. 
 \enqs

\ni (iii) The result is a consequence of (i) and (ii).\ep

\section{Example of a hybrid model}\label{sec:hybrid model}

In this section, we consider an explicit example where the random default time $\tau$ is given by a hybrid model as in \cite{CPS09, CL06} and the information flow $G$ is supposed to depend on the asset values at a horizon time which is similar to \cite{GJZ09}. 

Let $B=(B_t,t\geq 0)$ be a standard Brownian motion and $N^P=(N_t^P,t\geq 0)$ be a Poisson process with intensity $\lambda \in\mathbb R_+$. We suppose that $B$ and $N^P$ are independent. Let $\mathbb F=(\mathcal F_t)_{t\geq 0}$ be the complete and right-continuous filtration generated by the processes $B$ and $N^P$ where $\mathcal F_t=\cap_{s>t}\sigma\{B_u,N_u^P;u\leq s\}$. We define the default time $\tau$ by a hybrid model. More precisely,   consider a first asset process 
$S_t^1=\exp(\sigma B_t-\frac 12\sigma^2 t)$ where $\sigma>0$ and let $\tau_1=\inf\{t>0,S_t^1\leq l\}$ where $l$ is a given constant threshold such that $l<S_0^1$. 
In a similar way, consider a second asset process 
$S_t^2=\exp(\lambda t-N_t^P)$ and define $\tau_2=\inf\{t>0,N_t^P=1\}$. Let the default time be given by $\tau=\tau_1\wedge\tau_2$ which is an $\mathbb F$-stopping time with a predictable component $\tau_1$ and a totally inaccessible component $\tau_2$. Let the information flow $G$ be given by the vector $G=(S_{{T'}}^1, S_{{T'}}^2)$ where ${T'}>T$ is a horizon time. 

We first give the density of $G$ which is defined in \eqref{equ: jacod hypothesis}. By direct computations, 
\beqs
p_t(x_1,x_2) & = & \sqrt{\frac{T'}{T'-t}}\frac{\phi\Big(\frac{\ln(x_1)+\frac 12\sigma^2T'-\sigma B_t}{\sigma\sqrt{T'-t}}\Big)}{\phi\Big(\frac{\ln(x_1)+\frac 12\sigma^2T'}{\sigma\sqrt{T'}}\Big)}\cdot \\
& &  e^{\lambda t}\frac{(\lambda(T'-t))^{\lambda T'-\ln(x_2)-N_t^P}}{(\lambda T')^{\lambda T'-\ln(x_2)}}\frac{(\lambda T'-\ln(x_2))!}{(\lambda T'-\ln(x_2)-N_t^P)!}\indic_{\N}(\lambda T'-\ln(x_2)-N_t^P)
\enqs
where $\phi$ is the density function of the standard normal distribution $N(0,1)$, i.e. $\phi(x)=\frac{1}{\sqrt{2\pi}}e^{-\frac{x^2}{2}}$.  
Denote by $\tilde N^P$ the compensated Poisson process defined by
\beqs
\tilde N^P_t & =& N_t^P-\lambda t\;,\quad t\geq 0\;.
\enqs 
The dynamics of the assets processes are then given by
\beqs
dS^1_t & = &  S^1_t \sigma dB_t\;,\\
d S^2_t & = &  S^2_{t-}\Big((e^{-1}-1)d\tilde N_t^P+e^{-1}\lambda dt\Big)\;.
\enqs 
This leads to consider the driving martingale $M$ defined by $M=(\sigma B,(e^{-1}-1)\tilde N^P)^\top$.
Its oblique bracket of $M$ is then given by
$$
\langle M\rangle _t  = 
\left(
\begin{array}{cc}
\sigma^2 t  & 0\\
0 & (e^{-1}-1)^2\lambda t 
\end{array}
\right)\;,\quad t\geq 0 \;. 
$$
Then, we can write the dynamics of the asset processes using the notations of the previous section:
\beqs
d S^i_t & = & S^i_{t-}\Big(d M^i_t+\alpha^1_t d \langle M^i,M^1\rangle _t+\alpha^2_t d \langle M^i,M^2\rangle _t\Big)
\enqs
with 
\beqs
\alpha^1_t & = & 0 \\
\alpha^2_t & = & {e^{-1}\over (e^{-1}-1)^2}
\enqs
 for all $t\geq 0$.
We can then compute the terms $m^1$ and $m^2$ appearing in Lemma \ref{Lem mi} and we get
\beqs
m^1_t(x) & = & -{1\over\sigma\sqrt{T'-t}} {\phi'\over\phi}\Big(\frac{\ln(x_1)+\frac 12\sigma^2T'-\sigma B_t}{\sigma\sqrt{T'-t}}\Big) \\
 & = & \frac{\ln(x_1)+\frac 12\sigma^2T'-\sigma B_t}{\sigma^2(T'-t)}
 \enqs
 and
 \beqs
m^2_t(x) & = & \frac{1}{(e^{-1}-1)} \Big({\lambda T'-\ln(x_2)-N_{t-}^P\over \lambda (T'-t)}-1\Big)
\enqs
for $t\geq 0$. Since the matrix $\langle M\rangle $ is diagonal the  process $\mu$ given by Lemma \ref{lem mu} can be taken such that $\mu=(m^1,m^2)^\top$. We easily check that Assumption \ref{Hyp-mu} is satisfied. We can  then apply Theorem \ref{THM ORD INS} to the optimization problem with maturity $T$ and we get
\begin{itemize}
\item an optimal strategy for the ordinary agent given by
\beqs
\pi^{ord}_t  & = & \alpha_t \;,\quad t\in[0,T]\;, 
\enqs
and the maximal expected utility
\beqs
V^{\log}_\F & = & \log x +{1\over 2}\E\Big[\int_0^T\alpha_t^\top d \langle M\rangle _t\alpha_t\Big]~~=~~\log x +{e^{-1}\over (e^{-1}-1)^2}{\lambda T\over 4}\;,
\enqs
\item an optimal strategy for the insider given by
\beqs
\pi^{ins}_t  & = & \alpha_t +\indic_{\tau\leq t}\mu_t(G)\;,\quad t\in[0,T]\;, 
\enqs
and the maximal expected logarithmic utility
 \beqs
V^{\log}_\G 
 & = &\log x +{1\over 2}\E\Big[\int_0^T\alpha_t^\top d \langle M\rangle _t\alpha_t\Big]+{1\over 2}\E\Big[\int_0^T\mu_t(G)^\top d \langle M-M^\tau\rangle _t\mu_t(G)\Big]\\
 & = & \log x +{e^{-1}\over (e^{-1}-1)^2}{\lambda T\over 4}+
  {1\over 2}\E\Big[\int_0^T\indic_{t\geq\tau}\frac{\big(\ln(S^1_{T'})+\frac 12\sigma^2T'-\sigma B_t\big)^2}{\sigma^3(T'-t)^2}dt \Big] \\
  & & + {1\over 2}\E\Big[\int_0^T\indic_{t\geq\tau} \Big({\lambda T'-\ln(S^2_{T'})-N_{t}^P\over \lambda (T'-t)}-1\Big)^2\lambda dt \Big]\;, 
 \enqs
\item the insider's additional expected utility 
 \beqs
 V^{\log}_\G-V^{\log}_\F & = &  {1\over 2}\E\Big[\int_0^T\indic_{t\geq\tau}\frac{\big(\ln(S^1_{T'})+\frac 12\sigma^2T'-\sigma B_t\big)^2}{\sigma^3(T'-t)^2}dt \Big] \\
  & & + {1\over 2}\E\Big[\int_0^T\indic_{t\geq\tau} \Big({\lambda T'-\ln(S^2_{T'})-N_{t}^P\over \lambda (T'-t)}-1\Big)^2\lambda dt \Big]
\;
 \enqs
where
\[\begin{split}&\quad\E\Big[\int_0^T\indic_{t\geq\tau}\frac{\big(\ln(S^1_{T'})+\frac 12\sigma^2T'-\sigma B_t\big)^2}{\sigma^3(T'-t)^2}dt \Big]=\E\Big[\int_0^T\indic_{t\geq\tau}\frac{\big(B_{T'}-B_t\big)^2}{\sigma(T'-t)^2}dt \Big]\\
&=\mathbb E\Big[\int_{\tau\wedge T}^T\frac{1}{\sigma(T'-t)}dt\Big]=\sigma^{-1}\mathbb E\Big[\ln\Big(\frac{T'-\tau\wedge T}{T'-T}\Big)\Big]
\end{split}\]
and
\[
\begin{split}&\E\Big[\int_0^T\indic_{t\geq\tau} \Big({\lambda T'-\ln(S^2_{T'})-N_{t}^P\over \lambda (T'-t)}-1\Big)^2\lambda dt \Big]=\E\Big[\int_0^T\indic_{t\geq\tau} \Big({N_{T'}^P-N_{t}^P\over \lambda (T'-t)}-1\Big)^2\lambda dt \Big]\\
 & = \mathbb E\Big[\int_{\tau\wedge T}^T{dt\over T'-t}\Big]=\mathbb E\Big[\ln\Big(\frac{T'-\tau\wedge T}{T'-T}\Big)\Big]\;.
\end{split}\]
Hence we get
\beqs
V^{\log}_\G-V^{\log}_\F & = & (\sigma^{-1}+1)\mathbb E\Big[\ln\Big(\frac{T'-\tau\wedge T}{T'-T}\Big)\Big]\;.
\enqs
We note that the gain of the insider is strictly positive. In the limit case where $T'=T$, the insider may achieve a terminal wealth that is not bounded due to possible arbitrage strategies.  
\end{itemize}


\section{Conclusion}\label{sec:conclusion}

We study in this paper an optimal investment problem under default risk where related information is considered as an exogenous risk added at the default time.  The framework we present can also be easily adapted to information risk modelling for other sources of risks. The main contributions are twofold. First, the information flow is added at a random stopping time rather than at the initial time. Second,  we consider in the optimization problem a random time which does not necessarily satisfy the standard intensity nor density hypothesis in the credit risk. From the theoretical point of view, we  study the associated enlargement of filtrations and prove that Jacod's (H')-hypothesis holds in this setting. From the financial point of view, we obtain explicit logarithmic utility maximization results and compute the gain of the insider due to additional information.    
\bibliography{reference}
\bibliographystyle{plain}
\appendix
\section*{Appendix}
\subsection*{Proof of Proposition \ref{Pro:predictableprocess}}
\bproof
We begin with the proof of the ``if'' part. Assume that $Z$ can be written in the form \eqref{equ:Zpredictable} such that $Y$ is $\mathbb F$-predictable and $Y(\cdot)$ is $\mathcal P(\mathbb F)\otimes\mathcal E$-measurable. Since $\tau$ is an $\mathbb F$-stopping time, the stochastic interval $[\![0,\tau]\!]$ is a $\mathcal P(\mathbb F)$-measurable set. Hence the process $\indic_{[\![0,\tau]\!]}Y$ is $\mathbb F$-predictable and hence is $\mathbb G$-predictable. It remains to prove that the process $\indic_{]\!]\tau,+\infty[\![}Y(G)$ is $\mathbb G$-predictable. By a monotone class argument (see e.g. Dellacherie and Meyer \cite{DMI} Chapter I.19-24), we may assume that $Y(G)$ is of the form $Xf(G)$, where $X$ is a left-continuous $\mathbb F$-adapted process, and $f$ is a Borel function on $E$. Thus $\indic_{]\!]\tau,+\infty[\![}Xf(G)$ is a left-continuous $\mathbb G$-adapted process, hence is $\mathbb G$-predictable. Therefore, we obtain that the process $Z$ is $\mathbb G$-predictable.

In the following, we proceed with the proof of the ``only if'' part. Let $Z$ be a $\mathbb G$-preditable process. We first show that the process $Z\indic_{]\!]0,\tau]\!]}$ is an $\mathbb F$-predictable process. Again by a monotone class argument, we may assume that $Z$ is left continuous. In this case the process $Z\indic_{]\!]0,\tau]\!]}$ is also left continuous. Moreover, by the left continuity of $Z$ one has  \[Z_t\indic_{\{\tau\geq t\}}=\lim_{\varepsilon\rightarrow 0+}Z_{t-\varepsilon}\indic_{\{\tau> t-\varepsilon\}},\quad t>0.\]
Since each random variable $Z_{t-\varepsilon}\indic_{\{\tau> t-\varepsilon\}}$ is $\mathcal F_t$-measurable, we obtain that $Z_t\indic_{\{\tau\geq t>0\}}$ is also $\mathcal F_t$-measurable, so that the process $Y=Z\indic_{]\!]0,\tau]\!]}$ is $\mathbb F$-adapted and hence $\mathbb F$-predictable (since it is left continuous). Moreover, by definition one has $Z_t\indic_{\{\tau\geq t>0\}}=Y_t\indic_{\{\tau\geq t>0\}}$.

For the study of the process $Z$ on $\indic_{]\!]\tau,+\infty[\![}$, we use the following characterization of the predictable $\sigma$-algebra $\mathcal P(\mathbb G)$. The $\sigma$-algebra $\mathcal P(\mathbb G)$ is generated by sets of the form $B\times [0,+\infty)$ with $B\in\mathcal G_0$ and sets of the form $B'\times [s,s')$ with $0<s<s'<+\infty$ and $B'\in\mathcal G_{s-}:=\bigcup_{0\leq u<s}\mathcal G_u$. It suffices to show that, if $Z$ is the indicator function of such a set, then $\indic_{]\!]\tau,+\infty[\![}Z$ can be written as $\indic_{]\!]\tau,+\infty[\![}Y(G)$ with $Y(\cdot)$ being a $\mathcal P(\mathbb F)\otimes\mathcal E$-measurable function.

By \eqref{equ:filtrationG}, $\mathcal G_0$ is generated by $\mathcal F_0$ and sets of the form $A\cap\{\tau=0\}$, where $A\in\sigma(G)$. Clearly for any $B\in\mathcal F_0$, the function $\indic_{B\times [0,+\infty)}$ is already $\mathbb F$-predictable process. Let $U$ be a Borel subset of $E$ and $B=G^{-1}(U)\cap\{\tau=0\}$. Let $Y(\cdot)$ be the $\mathcal P(\mathbb F)\otimes\mathcal E$-measurable function sending $(\omega,t,x)\in\Omega\times\mathbb R_+\times E$ to $\indic_{\{\tau(\omega)=0\}}\indic_U(x)$. Then one has $\indic_{B\times[0,+\infty)}=Y(G)$. By a monotone class argument, we obtain that, if $Z$ is of the form $\indic_{B\times[0,+\infty)}$ with $B\in\mathcal G_0$, then there exists a $\mathcal P(\mathbb F)\otimes\mathcal E$-measurable function $Y(\cdot)$ such that $\indic_{]\!]\tau,+\infty[\![}Z=\indic_{]\!]\tau,+\infty[\![}Y(G)$.

In a similar way, let $s,s'\in(0,+\infty)$, $s<s'$. By \eqref{equ:filtrationG}, $\mathcal G_{s-}$ is generated by $\mathcal F_{s-}$ and sets of the form $A\cap\{\tau\leq u\}$ with $u<s$ and $A\in\sigma(G)$. If $B'\in\mathcal F_{s-}$, then the function $\indic_{B'\times [s,s')}$ is already an $\mathbb F$-predictable process. Let $U$ be a Borel subset of $E$ and $B'=G^{-1}(U)\cap\{\tau\leq u\}$. Let $Y(\cdot)$ be the $\mathcal P(\mathbb F)\otimes\mathcal E$-measurable function  sending $(\omega,t,x)\in\Omega\times\mathbb R_+\times E$ to $\indic_{\{\tau(\omega)\leq u\}}\indic_{[s,s')}(t)\indic_U(x)$, then one has $\indic_{B'\times[s,s')}=Y(G)$. Therefore, for any process $Z$ of the form $\indic_{B\times[s,s')}$ with $B\in\mathcal F_{s-}$, there exists a $\mathcal P(\mathbb F)\otimes\mathcal E$-measurable function $Y(\cdot)$ such that $\indic_{]\!]\tau,+\infty[\![}Z=\indic_{]\!]\tau,+\infty[\![} Y(G)$. The proposition is thus proved.
 
\finproof

\end{document}